\documentclass[preprint,12pt]{elsarticle}
\usepackage[pdfborder={0 0 0}]{hyperref} % from ther hyperlink 
\usepackage{amsthm,amsmath,amsfonts,latexsym,amssymb} %mathematic symbols 
\usepackage{graphics,fancyhdr,graphicx,subfigure} % for figures 
\usepackage[ruled,linesnumbered]{algorithm2e} % for algorithm insertion
\usepackage[margin=1in]{geometry}
\usepackage{multirow,booktabs} % consolidate table rows; table forms 
\usepackage[table]{xcolor}
\usepackage{listings}
\usepackage{tabularx}
\usepackage{placeins}
\usepackage{comment}
\usepackage{lineno}
\usepackage{lscape}
\usepackage{bm}
\usepackage{hyperref}
\usepackage{mathrsfs}
\usepackage{arydshln}
\usepackage{appendix}
\usepackage{float}
\usepackage{enumitem}
\usepackage{url}

\newcommand*{\review}{\textcolor{black}}
\definecolor{custom-blue}{RGB}{3,69,173}
\hypersetup{
    colorlinks = true,
    urlcolor   = blue,
    citecolor  = custom-blue,
}

\biboptions{numbers,sort&compress} % for refenrece format 
\definecolor{listinggray}{gray}{0.9}
\definecolor{lbcolor}{rgb}{0.9,0.9,0.9}
\definecolor{Darkgreen}{RGB}{0,100,0}

\begin{document} % for all document 
\abovedisplayskip=6.0pt
\belowdisplayskip=6.0pt
\begin{frontmatter} % for the preface and abstract 

\title{Causality-Respecting Adaptive Refinement for PINNs: Enabling Precise Interface Evolution in Phase Field Modeling}

\author[1,3]{Wei Wang}
\ead{jacques-wei.wang@connect.polyu.hk}
\author[2]{Tang Paai Wong}
\ead{dillian.wong@polyu.edu.hk}
\author[1]{Haihui Ruan \corref{cor1}}
\ead{hhruan@polyu.edu.hk}
\author[3]{Somdatta Goswami\corref{cor1}}
\ead{somdatta@jhu.edu}

\address[1]{Department of Mechanical Engineering, The Hong Kong Polytechnic University, Hong Kong, China}
\address[2]{University Research Faculty in Big Data Analytics, The Hong Kong Polytechnic University, Hong Kong, China}
\address[3]{Department of Civil and Systems Engineering, Johns Hopkins University, Baltimore, USA}
\cortext[cor1]{Corresponding author.}

\begin{abstract}
\noindent 
Physics-informed neural networks (PINNs) have emerged as a powerful tool for solving physical systems described by partial differential equations (PDEs). However, their accuracy in dynamical systems, particularly those involving sharp moving boundaries with complex initial morphologies, remains a challenge. This study introduces an approach combining residual-based adaptive refinement (RBAR) with causality-informed training to enhance the performance of PINNs in solving spatio-temporal PDEs. Our method employs a three-step iterative process: initial causality-based training, RBAR-guided domain refinement, and subsequent causality training on the refined mesh. Applied to the Allen-Cahn equation, a widely-used model in phase field simulations, our approach demonstrates significant improvements in solution accuracy and computational efficiency over traditional PINNs. Notably, we observe an `overshoot and relocate' phenomenon in dynamic cases with complex morphologies, showcasing the method's adaptive error correction capabilities. This synergistic interaction between RBAR and causality training enables accurate capture of interface evolution, even in challenging scenarios where traditional PINNs fail. Our framework not only resolves the limitations of uniform refinement strategies but also provides a generalizable methodology for solving a broad range of spatio-temporal PDEs. \review{The enhanced performance of the RBAR–causality combined framework demonstrates its strong potential for advancing PINN-based modeling of physical systems characterized by complex, evolving interfaces.}
\begin{keyword}
physics-informed neural network \sep residual-based adaptive refinement \sep causality training \sep phase field modeling \sep Allen-Cahn equations  
\end{keyword}

\end{abstract}
\end{frontmatter}

\section{Introduction}
\label{sec:intro}

Partial differential equations (PDEs) are fundamental in science and engineering, modeling phenomena from fluid flow to material failure. While analytical solutions are limited, numerical methods have been the primary approach for solving PDEs. Traditional physics-based numerical methods like finite element, isogeometric analysis, and spectral methods offer high accuracy but can be computationally expensive. In recent years, deep learning (DL) has emerged as a promising alternative for fast predictions of physical systems.

Within the broader category of DL techniques, physics-informed neural networks (PINNs) \cite{raissi2019physics,samaniego2020energy} have gained particular attention as an alternative to traditional PDE solvers. PINNs approximate PDE solutions by training a deep neural network (DNN) to minimize a loss function that incorporates initial and boundary conditions, as well as the PDE residual at selected points (collocation points). This approach is essentially a mesh-free technique that converts the problem of directly solving governing equations into a loss function optimization problem. PINNs integrate the mathematical model into the network without requiring labeled data, reinforcing the loss function with a residual term from the governing equation that acts as a penalizing term, thus ensuring the network considers the underlying physics rather than solely fitting data. Numerous variants of PINNs have been developed including conservative PINNs \cite{jagtap2020conservative}, sparse PINNs \cite{ramabathiran2021spinn}, variational PINNs \cite{goswami2020transfer}, domain-decomposition PINNs (XPINNs) \cite{jagtap2020extended}, and separable PINNs \cite{cho2022separable}, stochastic PINNs \cite{karumuri2020simulator} to name a few. However, PINNs are not the only methods that constrain ML models with physics. Other popular methods that use DNNs to solve PDEs include the Deep Ritz method \cite{yu2018deep} and approaches based on the Galerkin or Petrov-Galerkin method \cite{kharazmi2019variational} including the Deep Galerkin Method \cite{sirignano2018dgm}. When a Galerkin approach is used on collocation points, the framework is a variant of PINNs, \textit{i.e.} hp-VPINNs \cite{kharazmi2019variational}. In the Deep Energy Method \cite{samaniego2020energy} the loss function minimizes the variational energy of the system at quadrature points. Other methods have used Gaussian Processes (GPs) to solve PDEs \cite{Yang2018,Pang2020} or have imposed physical constraints on GPs \cite{Swiler2020}.

While PINNs have demonstrated success across various domains, their application to phase field equations has been limited. This scarcity in literature stems from the challenges posed by sharp transition interface in phase field solutions, which evolve over time \cite{wight2020solving}. Consequently, certain aspects of phase field model solutions prove more challenging to learn, both spatially and temporally. To address these spatio-temporal challenges, researchers have recently implemented sequential training strategies \cite{krishnapriyan2021characterizing, mattey2022novel}. The work in \cite{wang2022respecting} has shown that these strategies are rooted in the principle of causality, suggests that temporal evolution in dynamic systems may lead PINNs to converge towards erroneous solutions \cite{mattey2022novel}. While a straightforward causality-based training method can help PINNs avoid these erroneous solutions, their predictive capability in forecasting sharp moving layers remains insufficient \cite{wight2020solving}. Accurate capture of complex morphology in moving layers necessitates adaptive sampling of residual points over the computational domain. Residual-based adaptive refinement (RBAR) methods have been proposed, adding points in areas with large residual losses following different sampling rules \cite{yu2022gradient,wu2023comprehensive}. However, these refinement methods treat spatial and temporal dimensions equivalently, disregarding the significance of causality.

In this study, we propose a combined RBAR-causality method to solve the Allen-Cahn equations with high accuracy. To the best of our knowledge, this work presents the first integration of causality-based training and residual-based adaptive refinement within physics-informed neural networks (PINNs) to achieve accurate surrogate modeling of solutions with sharp interfaces. We demonstrate the effectiveness of the proposed approach using the Allen–Cahn equation. Our specific contributions are: \vspace{-10pt}
\begin{itemize}
\item We demonstrate that our straightforward approach can effectively resolve the Allen-Cahn equation, characterized by a complex sharp-moving interface. This eliminates the need for repetitive training cycles or time-step segmentation of the network, as seen in previous studies \cite{wight2020solving,jung2024ceens}. The simplicity of our proposed RBAR-causality method facilitates its potential generalization across various spatiotemporal systems. \vspace{-10pt}
\item We identify and discuss the `overshoot and relocate' phenomenon observed in the RBAR-causality combined PINN. This phenomenon underscores the exceptional collaborative efficiency of RBAR and causality, highlighting the adaptive error correction capability of the combined approach. \vspace{-10pt}
\end{itemize}

\section{Enhanced PINNs: Integrating Residual-Based Adaptive Refinement and Causality Training}
\label{sec:methods}
\noindent In this section, we introduce the PINNs framework followed by our proposed residual-based adaptive refinement causality training framework. Consider an initial-boundary value problem defined by the following equations:
\begin{align}
\mathcal N[{u(\bm{x},t)}](\bm{x}) & = f(\bm{x}, t), \bm{x} \in {\Omega}, t \in (0, T] \label{PDE_f}\\
\mathcal B[{u(\bm{x},t)}](\bm{x}) & = g(\bm{x},t),\bm{x} \in {\partial{\Omega}}, t \in (0, T] \label{PDE_BC}\\
u(\bm{x},0) & = h(\bm{x}), \bm{x} \in \Omega \backslash \partial{\Omega} \label{eq:PDE_IC}
\end{align}
where $\Omega \subset \mathbb{R}^d$ denotes a bounded domain on which the PDE is defined, $\partial\Omega$ denotes the domain boundary, $u: \Omega \times [0, T] \to \mathbb{R}$ is the desired solution, $\boldsymbol{x} \in \Omega$ is a spatial vector variable, $t$ is time, and $\mathcal{N}$ and $\mathcal{B}$ are spatio-temporal differential operators. In \autoref{eq:PDE_IC}, $f: \Omega \to R$ is the driving function, $g: \partial{\Omega} \to R$ is the boundary condition function, and $h: \Omega \backslash  \partial{\Omega} \to R$ denotes the initial condition. In PINN, the assumed parametric form of the solution is constructed using DNN as a global space-time function approximator. Specifically, the solution $u(\bm{x}, t)$ is approximated as:
\begin{equation}
u(\bm{x},t) \approx u_{NN}(\bm{x}, t; \boldsymbol{\theta}).
\end{equation}
where $\boldsymbol{\theta}$ denotes all trainable network parameters that globally define the solution over the entire space–time domain. The neural network representing $u_{NN}(\bm{x}, t; \bm{\theta})$ is a fully connected feed-forward network with five hidden layers, each containing 100 neurons and using the hyperbolic tangent activation function. The network weights are initialized using the Xavier Normal method.

To satisfy the conditions expressed in Eqn.\eqref{PDE_f}--\eqref{eq:PDE_IC}, the PINNs model is trained by minimizing the following composite loss function, $\mathcal L(\boldsymbol{\theta},\beta_b,\beta_i)$, defined as:
\begin{equation}
    \mathcal L(\boldsymbol{\theta},\beta_b,\beta_i) = \mathcal L_r(\boldsymbol{\theta})  + \beta_{b}\mathcal L_{b}(\boldsymbol{\theta})+ \beta_i\mathcal L_i(\boldsymbol{\theta}),
\label{loss_fun}
\end{equation}
where $\mathcal{L}_r$, $\mathcal{L}_b$, and $\mathcal{L}_i$ are loss terms derived from the PDE itself (Eq. \eqref{PDE_f}), the boundary condition (Eq. \eqref{PDE_BC}), and the initial condition (Eq.\eqref{eq:PDE_IC}), respectively. The hyperparameters  $\beta_b$ and $\beta_i$ provide flexibility in assigning different weights to each term, facilitating their interaction during the training process. These weights can be either user-specified or automatically tuned during training \cite{ wang2021understanding,wang2022and}. In this work, we determine the values of these weights using the self-adaptive method \cite{mcclenny2020self}. The individual loss terms are defined as follows:
\begin{align}
\mathcal L_r(\boldsymbol{\theta}) & = \frac{1}{N_r}\sum_{i=1}^{N_r}(\mathcal{N}[u(\bm{x}_r^i,t_r^i; \bm{\theta})] - f(\bm{x}_r^i,t_r^i))^2,\\
\mathcal L_b(\boldsymbol{\theta}) & = \frac{1}{N_b}\sum_{i=1}^{N_b} (\mathcal{B}[u(\bm{x}_b^i,t_b^i; \bm{\theta})] - g(\bm{x}_b^i,t_b^i))^2, \label{loss_b} \\
\mathcal L_i(\boldsymbol{\theta}) & = \frac{1}{N_i}\sum_{i=1}^{N_0} (u(\bm{x}_0^i,0; \bm{\theta}) - h(\bm{x}_0^i))^2, \label{loss_i}
\end{align}
where $N_r$, $N_b$, and $N_i$ denote the number of collocation points, boundary points and initial condition points, respectively. The training points the respective set is denoted as: $\left\{\bm{x}_r^i,t_r^i\right\}_{i=1}^{N_r}$, $\left\{\bm{x}_b^i,t_b^i\right\}_{i=1}^{N_b}$, and $\left\{\bm{x}_0^i\right\}_{i=1}^{N_0}$.  

In this work, the $L^2$-norm (mean squared error) is adopted for all the loss functions $\mathcal{L}_* $ ($* \in \{r, b, i\}$) . This choice is motivated not only by its differentiability and optimization-friendly properties but also by its physical interpretation: minimizing the $L^2$-norm of the PDE residual is analogous to minimizing the ‘energy’ of the violation of the physical laws across the computational domain. This strongly penalizes large deviations from the governing equations, thereby enforcing the physical constraints more effectively. 

\subsection{Causality training method}
To mitigate the issue of the erroneous convergence of PINN caused by the temporal evolution in spatio-temporal equations, the causality training method is proposed by defining temporal residual loss function $\mathcal L_r(t,\bm\theta)$ with spatial set $\left\{\bm{x}^j_r \right\}_{j=1}^{N_x}$ \cite {wang2022respecting}:
\begin{equation}
    \mathcal{L}_r(t, \bm{\theta}) = \frac{1}{N_x}\sum_{j=1}^{N_x}(\mathcal{N}[u(\bm{x}_r^j,t; \bm{\theta})] - f(\bm{x}_r^j,t))^2,
\end{equation}
where $t$ denotes time, and the total time is divided into $N_t$ intervals ($\left\{ t_k\right\}_{k=1}^{N_t}$). Thus, the reformulation of the residual loss function is given by:
 \begin{equation}
\mathcal L_r(\boldsymbol{\theta}) = \frac{1}{N_t}\sum_{k=1}^{N_t} \omega_{k}\mathcal L_r(t_k,\bm\theta),  \label{loss_r_cau}
\end{equation}
the residual loss of each interval is weighted by $\omega_k$, expressed as:
\begin{equation}
\omega_{k} = exp\left[-\epsilon \sum_{m=1}^{k-1}\mathcal{L}_r(t_m, \bm{\theta})\right],\text{ for } k = 2,3,...N_t.
\label{weight_cau}
\end{equation}
The causality parameter, $\epsilon =10$, in this work, controls the steepness of the weights $\omega_k$. These weights are inversely and exponentially proportional to the cumulative residual loss from previous time steps. As a result, $\mathcal{L}_r(t_k,\bm{\theta})$ will not be minimized unless all preceding residual losses ${\mathcal{L}_r(t_m,\bm{\theta})}_{m=1}^{k-1}$ are reduced below a predetermined critical threshold. This reformulation ensures that the PINN is well-trained at every time step, effectively incorporating the principle of causality into the training process.

\subsection{Residual-based adaptive refinement}
\label{subsec:res_adaptive}

\begin{figure}[H]
\begin{center}
\includegraphics[width=0.7\textwidth]{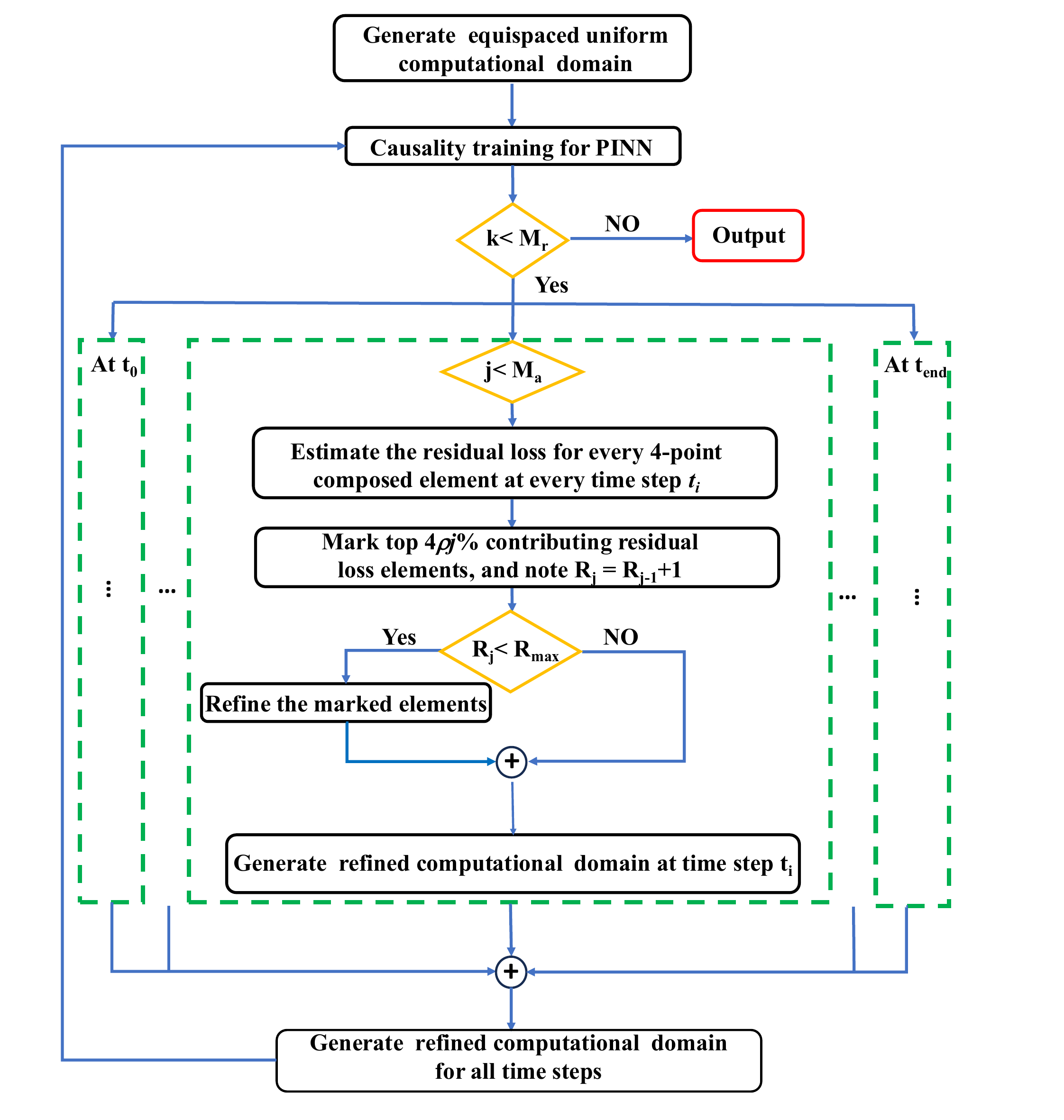}
\caption{Conceptual workflow of a mesh-free RBAR–PINN framework using causality training and residual-based adaptive refinement.}
\label{Fig:Flowchart}
\end{center}
\end{figure}

To achieve a more accurate solution, we propose a Residual-Based Adaptive Refinement (RBAR) method that dynamically refines areas with large residual loss at each time step. Figure \ref{Fig:Flowchart} illustrates the steps in our framework. Our adaptive h-refinement scheme, which discretizes the problem domain into multiple elements to achieve higher accuracy, is inspired by Goswami \textit{et al}. \cite{goswami2020adaptive}, while the residual-based refinement approach is influenced by Wu \textit{et al}. \cite{wu2023comprehensive}. In this mesh-free context, the domain is partitioned into `elements' solely for the purpose of localized residual evaluation and ranking. To simplify the description, this conceptual region is denoted as `element' the in methodology section. In the context of causality, we apply refinement exclusively to the spatial dimension.
The RBAR method identifies areas of the computational domain with the largest residual loss at each time step. This refinement typically occurs at sharp interfaces to enhance optimization. The RBAR scheme comprises three main steps:\vspace{-8pt}
\begin{enumerate}[leftmargin=*]
\item \textbf{Estimate}:
\begin{itemize}[leftmargin=*, nosep]
\item Initialize residual points as nodes of equispaced uniform grids within the computational domain.
\item After initial causality training, determine the residual loss at each time step for every point.
\item Partition the computational domain into `elements' , each composed of four adjacent points.
\item Calculate the residual loss of element as the sum of its constituent points' losses.
\end{itemize}
\item \textbf{Mark}: 
\begin{itemize}[leftmargin=*, nosep]
    \item Sort `elements' in descending order of their residual loss.
    \item At each adaptive refinement cycle $j$, the top $4\rho j\%$ of the originally discretized elements, corresponding to the largest residual errors, are marked for refinement, where $\rho = 5$ is a fixed ratio parameter and $j \in \{1,2,\ldots,M_a\}$ with $M_a = 5$ denoting the maximum number of adaptive refinements per time step.
    \item Ensure `elements' do not exceed a predefined maximum refinement level ($R_\text{max}$).
    \item Record the refinement level of each marked `element'.
\end{itemize}
\item \textbf{Refine}: 
\begin{itemize}[leftmargin=*, nosep]
    \item Apply h-refinement to marked `elements'.
    \item Add new points at the midpoint of each edge and at the center of the marked `element'.
    \item Replace the original `element' with four new `elements'.
    \item Repeat this refinement process 5 times across the computational domain.
\end{itemize}
\end{enumerate}

\noindent We propose the following algorithm for implementing the Residual-Based Adaptive Refinement (RBAR) method in the context of causality:

\begin{algorithm}[H]
\SetAlgoLined
\caption{RBAR–Causality Training Procedure}
\KwIn{
    $\Omega$: Initial computational domain (collocation elements)\\
    $T = \{t_1,\ldots,t_{N_t}\}$: Discrete time instances\\
    $M_r$: Maximum global refinement iterations\\
    $M_a$: Maximum adaptive refinements per time step\\
    $\rho$: Refinement ratio parameter\\
    $R_{\max}$: Maximum refinement level
}
\KwOut{$\mathcal{M}$: Trained PINN model, $\Omega_f$: Final refined domain}
Initialize PINN model $\mathcal{M}$ with parameters $\boldsymbol{\theta}$\;
Set initial domain $\Omega_0 \leftarrow \Omega$\;
\textbf{Initial training:}\\
Train $\mathcal{M}$ on $\Omega_0$ using causality-based loss minimization\;
\For{$k \leftarrow 1$ \KwTo $M_r$}{
    $\Omega_k \leftarrow \Omega_{k-1}$\;
    \ForEach{$t \in T$}{
        \For{$j \leftarrow 1$ \KwTo $M_a$}{
            Compute residual loss $\mathcal{L}_r(\bm{x},t)$ for all elements in $\Omega_k$\;
            Sort elements in descending order of $\mathcal{L}_r$\;
            Mark the top $4\rho j\%$ of the originally discretized elements for refinement\;
            \ForEach{marked element}{
                Increase refinement level $R \leftarrow R+1$\;
                \If{$R \le R_{\max}$}{
                    Apply $h$-refinement to the element\;
                }
            }
        }
    }
    \textbf{Causality training:}\\
    Retrain $\mathcal{M}$ on the refined domain $\Omega_k$ using causality-weighted temporal loss\;
}
Set $\Omega_f \leftarrow \Omega_k$\;
\Return{$\mathcal{M}, \Omega_f$}\;
\end{algorithm}

In each causality training cycle, we implement an adaptive learning rate scheduler based on the Stochastic Gradient Descent with Warm Restarts method proposed by Loshchilov and Hutter \cite{loshchilov2016sgdr}. The learning rate, $l_r$, is dynamically adjusted using a cosine decay schedule: it starts at $l_r = 5\times 10^{-3}$, gradually decreases to a predefined minimum or zero, and then restarts in successive cycles to facilitate stable training. This cyclical strategy offers several advantages: it facilitates efficient exploration of the loss landscape in the early stages of each cycle, allows for fine-tuning and convergence to optimal value as the rate decreases, and helps the model escape suboptimal local minima through periodic restarts. By employing this technique, we significantly enhance the robustness of our RBAR-causality based PINNs framework, effectively mitigating the risk of convergence to erroneous solutions that may arise from consistently high learning rates, while also avoiding the premature convergence often associated with monotonically decreasing rates. Furthermore, this approach aligns well with the iterative nature of our RBAR-Causality method, as each refinement stage can benefit from a fresh cycle of the learning rate schedule, potentially leading to improved adaptation to the refined mesh structure. 

\section{Phase Field modeling employing Allen-Cahn equation}
\label{sec:phase_field}

In this section, we derive a phase field model (PFM) based on the Allen-Cahn equation in both static and dynamic forms. We construct a two-phase PFM with a single order parameter $u$, beginning with the total Helmholtz free energy of the investigated systems:
\begin{equation}
\mathcal F = \int_\Omega f d\omega = \int_\Omega(W g(u) + \frac{\kappa}{2}|\nabla u|^2)d\omega.
\label{Helmholtz}
\end{equation}
Here, $W$ represents the energy barrier and $\kappa$ denotes the scale factor of the interfacial energy density. The function $g(u)=u^2(1-u)^2$ is the double well function, a fundamental component in phase field modeling \cite{lin2019phase, moelans2008introduction}. To describe the phase evolution, we employ the functional derivative of the Helmholtz free energy. Consequently, the corresponding Allen-Cahn equation for this system is given by:
\begin{equation}
\frac{\partial u}{\partial t} = -L_\sigma\frac{\delta \mathcal F}{\delta u}= -L_\sigma(2Wu(1-u)(1-2u)-\kappa \nabla^2u),
\label{AC_static}
\end{equation}
where $L_\sigma$ denotes the relaxation coefficient, which is proportional to the rate of phase evolution \cite{fan1997computer}. It is important to note that Eq. \eqref{AC_static} describes a static co-existence of two phases with a finite-thickness interface \cite{kim1999phase}.
To extend the model to dynamic systems with evolving phases, we introduce an additional driving force term to the static PFM:
\begin{equation}
\begin{aligned}
\frac{\partial u}{\partial t} & = -L_\sigma\left[2Wu(1-u)(1-2u)-\kappa \nabla^2u\right] + L_\eta\left[\frac{dP(u)}{du} f(u,\bm x,t)\right]\ \\
&= -L_\sigma\left[2Wu(1-u)(1-2u)-\kappa \nabla^2u\right] + L_\eta\left[\frac{dP(u)}{du}c\right].
\label{AC_dynamic}
\end{aligned}
\end{equation}
In this extended form, $L_\eta$ represents the scaling coefficient for the driving force term, and $f(u,\bm x,t)$ is the driving force function depending on the order parameter $u$, spatial coordinates $\bm x$, and time $t$. For simplicity in this study, we assume $f(u,\bm x,t)$ to be a constant $c$, representing a constant driving force. The term $\frac{dP(u)}{du} = u^3(10 - 15u +6u^2)$ ensures that the driving force originates from the interface, as proposed by Wang \textit{et al}. \cite{wang1993thermodynamically}. This formulation provides a comprehensive framework for modeling phase evolution in both static and dynamic contexts, capturing the interplay between interfacial energy, bulk free energy, and external driving forces.

\section{Numerical results}

In this section, we employ the proposed RBAR causality-based PINN in both static and dynamic cases with different initial morphology, \textit{i.e.}, plane interface and hump interface, respectively. The static and the dynamic cases are denoted with Eq. \eqref{AC_static} and \eqref{AC_dynamic}, respectively. The solutions obtained via our PINN framework are systematically benchmarked against the FEM solution computed using COMSOL Multiphysics. \review{Through comprehensive comparative analysis, we demonstrate the computational efficiency, numerical accuracy, and adaptive error correction capabilities of the RBAR causality-based PINN methodology in comparison with the conventional PINN schemes.}

\subsection{Quasi-static Allen-Cahn Equation}
\begin{figure}[H]
\begin{center}
\includegraphics[width=0.9\textwidth]{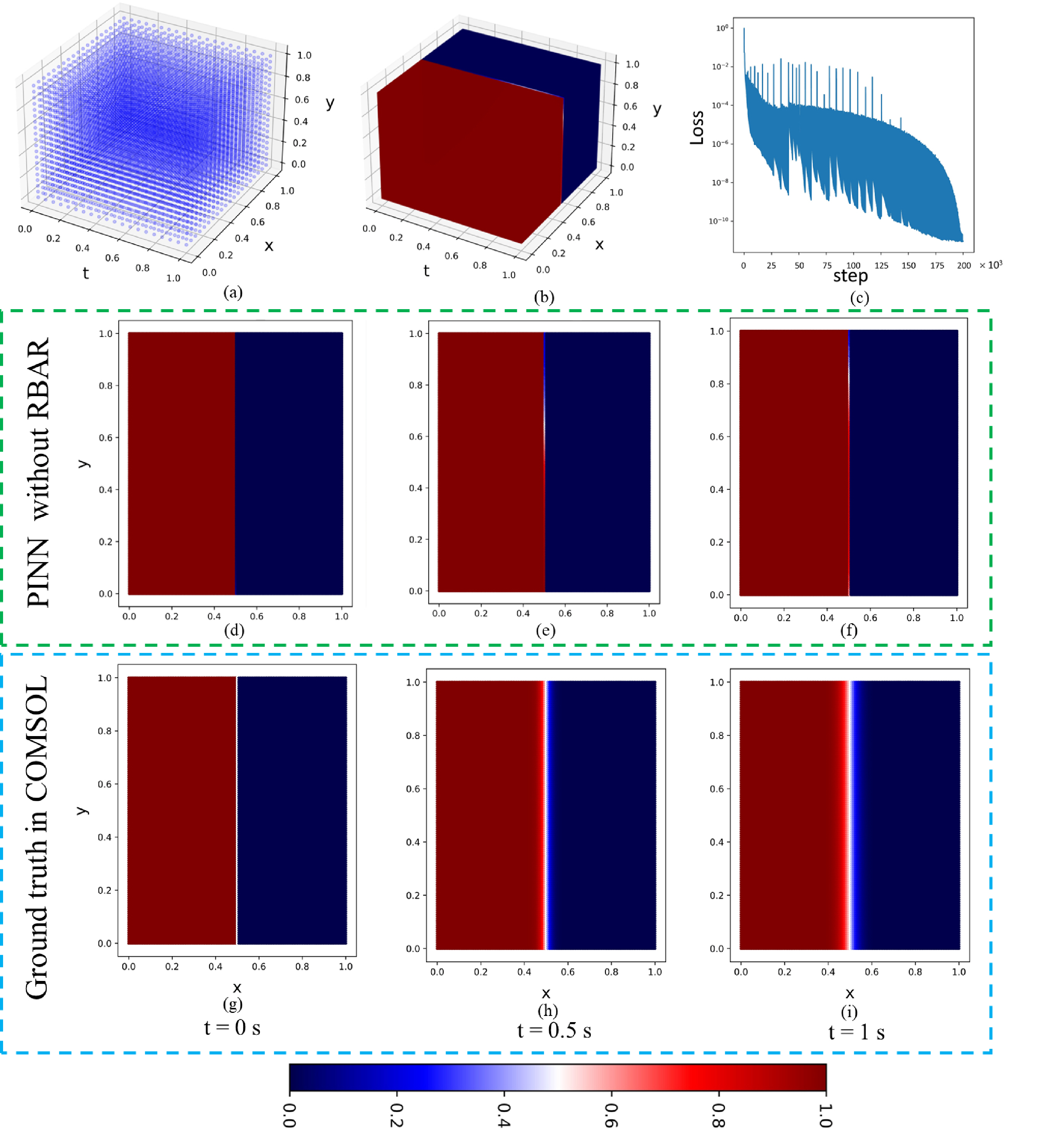}
\caption{Static phase evolution with planar interface using conventional PINN. (a) Initial computational domain discretized with 8,000 ($20 \times 20 \times 20$) uniformly distributed collocation points. (b) Three-dimensional evolution of the phase morphology simulated using the PINN framework. (c) History of the mean training loss obtained with the Adam optimizer over 20,000 iterations. (d–f) Phase morphology evolution predicted by \underline{PINN without RBAR refinement} at $t = 0$, $0.5$, and $1$ s, respectively. (g–i) Corresponding phase morphology evolution computed using COMSOL FEM at the same time points for direct comparison. The color scale represents the phase parameter $u$, with red ($u = 1$) and blue ($u = 0$) denoting distinct phases.}
\label{Fig.2_static_plane}
\end{center}
\end{figure}

We conduct our numerical experiments in a three-dimensional spatio-temporal domain where $(t,x,y) \in [0,1]\times [0,1] \times [0,1]$. The initial discretization comprises 8000 uniformly distributed collocation points prior to refinement, as depicted in Figure \ref{Fig.2_static_plane}(a). Our two-phase PFM simulation tracks the phase transition dynamics, with the full three-dimensional evolution shown in Figure \ref{Fig.2_static_plane}(b). The static Allen-Cahn equation takes the form in Eq. \eqref{AC_static}. And  we set the corresponding parameters: $L_\sigma = 1 \; m^3(J\cdot s)^{-1}$, $W = 8 \times 10^{-4} \; J\cdot  m^{-3}$, and $\kappa = 1 \times 10^{-4} \; J\cdot m^{-1}$. The initial condition, illustrated in Figure 1(d), establishes a discrete plane interface at $x = 0.5$ separating two phases: the red phase ($u = 1$) and the blue phase ($u = 0$). The following observations were made in the initial implementation of the framework, which employs only RBAR (without causality training):
\begin{enumerate}[leftmargin=*]
    \item The mean loss using the Adam optimizer \cite{kingma2014adam} converges to below $10^{-10}$ (see Figure \ref{Fig.2_static_plane}(c)), yet significant discrepancies emerge between PINN predictions (see Figures \ref{Fig.2_static_plane}(d-f)) and COMSOL FEM results (see Figures \ref{Fig.2_static_plane}(g-i)).
    \item While the interface evolves toward a finite width configuration (see Figure \ref{Fig.2_static_plane}(g-i)), the current distribution of collocation points proves insufficient to accurately capture the interface width variations. 
\end{enumerate}
These limitations highlight the necessity for implementing the complete RBAR strategy to improve solution accuracy, particularly in regions with steep gradients near the phase interface.

\begin{figure}[H]
\begin{center}
\includegraphics[width=1\textwidth]{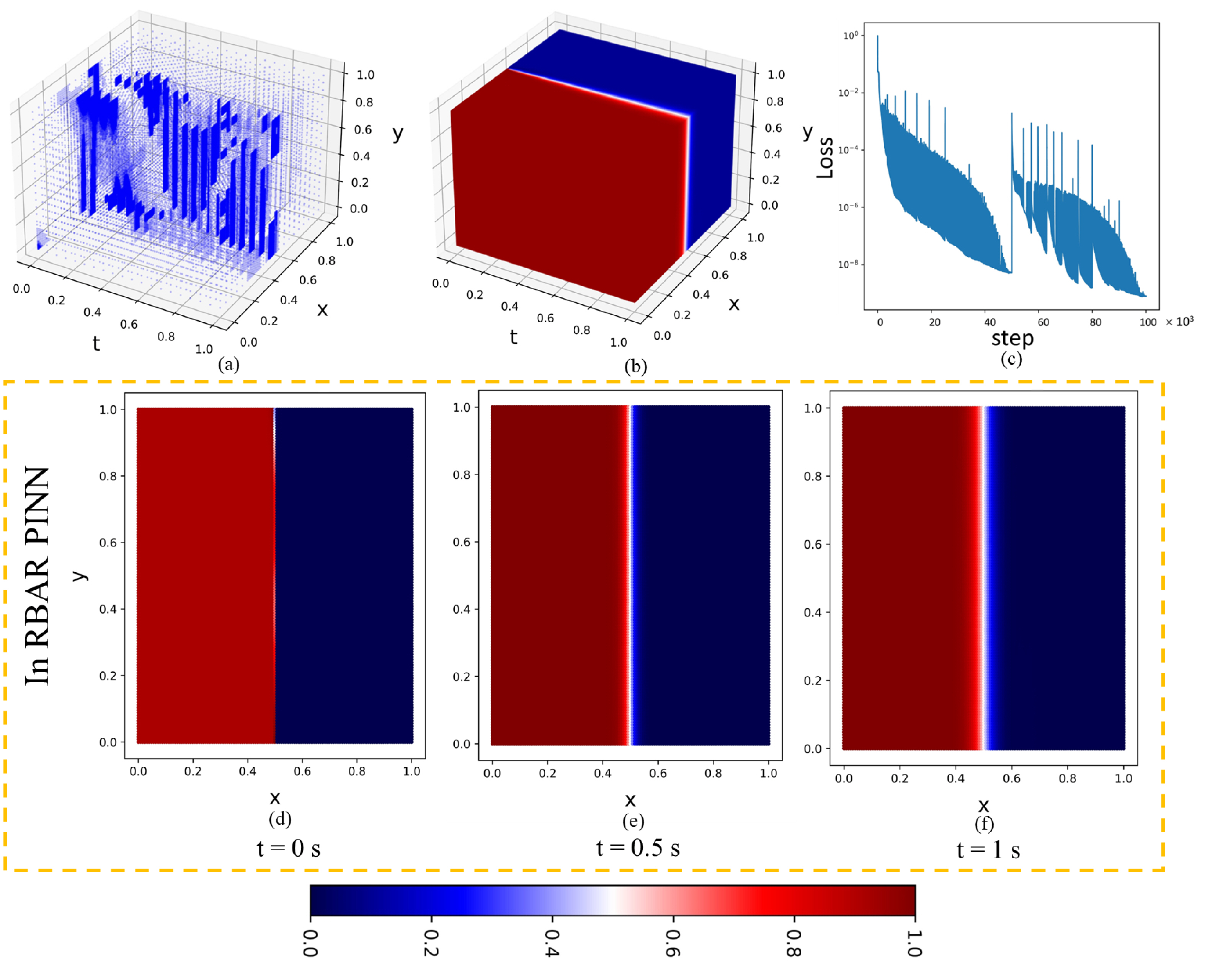}
\caption{Static phase evolution with planar interface using the \underline{RBAR-PINN} methodology. (a) Computational domain after RBAR refinement, showing the adaptive distribution of collocation points. (b) Three-dimensional evolution of the phase morphology predicted by RBAR-PINN. (c) Mean training loss per collocation point using the Adam optimizer over 100,000 iterations, including 50,000 pre-refinement and 50,000 post-RBAR iterations. (d–f) Phase morphology evolution captured by RBAR-PINN at $t = 0$, $0.5$, and $1$ s, illustrating the interface dynamics.}
\label{Fig.2_static_plane_RBAR}
\end{center}
\end{figure}

Following a single iteration of the computational domain with the collocation points in Figure \ref{Fig.2_static_plane}(a), an adaptive refinement was implemented employing the RBAR method, as shown in Figure \ref{Fig.2_static_plane_RBAR}(a). The refinement process automatically identified regions of highest residual loss, which predominantly occurred along the sharp phase interface. While some of the refined points appear to be randomly distributed, resulting in an irregular and seemingly counterintuitive pattern, this behavior stems from the inherently non-linear optimization process of PINNs. Factors such as training dynamics, sensitivity to initialization, and convergence to local minima can introduce significant errors in unexpected regions of the domain, thereby leading to a degree of randomness in the refinement pattern. This adaptive point distribution proved crucial for several reasons:
\begin{enumerate}[leftmargin=*]
    \item Enhanced Resolution: The increased density of collocation points near the interface enabled accurate capture of the steep gradients characteristic of phase boundaries.
    \item Improved Physics Representation: The refined mesh successfully resolved the interface widening phenomenon, as evidenced in Figure \ref{Fig.2_static_plane_RBAR}(d-f). The simulation results now demonstrate excellent agreement with the expected physical behavior.
    \item Computational Efficiency: By concentrating computational resources in regions of high physical activity (i.e., the phase interface), RBAR optimizes the distribution of collocation points to maximize accuracy while minimizing computational overhead.
\end{enumerate}
The effectiveness of RBAR lies in its ability to identify and adaptively refine regions where the PDE residuals are largest, thereby providing the PINN with the necessary spatial resolution to accurately characterize the underlying physics of the Allen-Cahn equation. This targeted refinement strategy proves particularly valuable for problems featuring localized phenomena such as moving interfaces and sharp gradients. The loss history with the RBAR PINNs framework is shown in Figure \ref{Fig.2_static_plane_RBAR}(c).

\subsection{Dynamic Allen-Cahn Equation}

\begin{figure}[H]
\begin{center}
\includegraphics[width=1\textwidth]{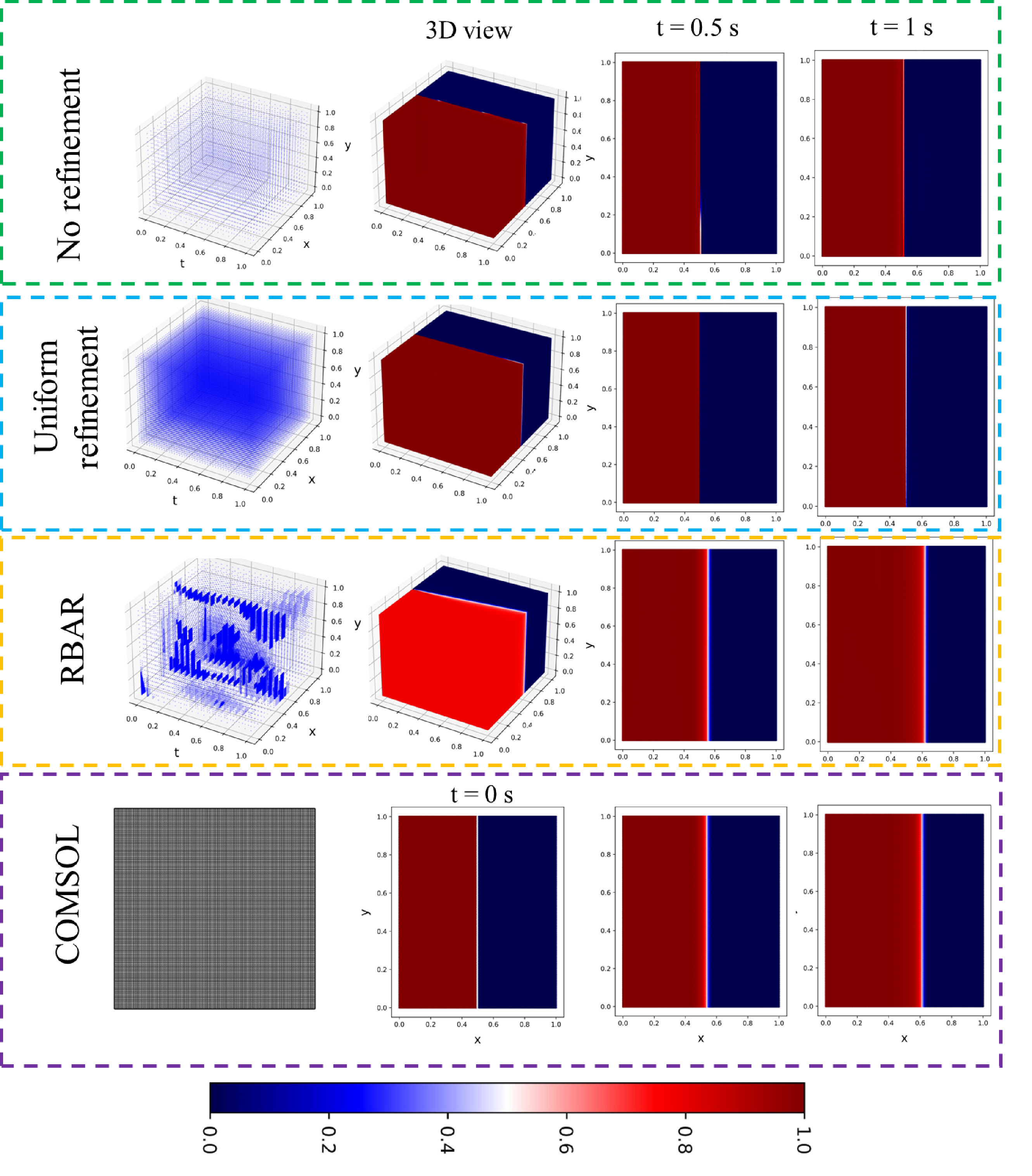}
\caption{Comparative analysis of dynamic phase evolution with planar interface using different computational approaches: unrefined PINN simulation showing static interface behavior, uniform mesh refinement ($44^3$ collocation points) demonstrating persistent stagnation, RBAR-enhanced PINN capturing interface motion, and reference COMSOL FEM solution validating the dynamic evolution.}
\label{Fig.3_dynamic_plane}
\end{center}
\end{figure}

For the dynamic case analysis, we maintain the same computational domain configuration as established in the static case. The governing Allen-Cahn equation incorporates a constant driving force term as follows:
\begin{equation}
\frac{\partial u}{\partial t} = W\left[2u(1-u)(1-2u)\right] - \kappa\nabla^2u + L_{\eta}c u^3(10-15u+6u^2),
\label{AC_dynamic_n}
\end{equation}
where $W = 8\times 10^{-4} \; J\cdot m^{-3}$, $\kappa = 1\times 10^{-4} \; J\cdot m^{-1}$, $L_{\eta} = 0.01 \; s^{-1}$, and $c = 200$. The initial condition consists of a planar interface morphology. In this investigation of moving plane dynamics, we implement RBAR PINN without the causality training strategy.
Impact of Refinement Strategies. Our observations are summarized as follows:
\begin{enumerate}
    \item Without Refinement: As shown in Figure \ref{Fig.3_dynamic_plane}, the planar interface remains stationary, failing to capture the expected phase transition dynamics.
    \item Uniform Refinement: Even after increasing the resolution from $20^3$ to $44^3$ collocation points (exceeding 8,100 total points), the interface exhibited no displacement despite 100,000 training iterations.
\end{enumerate}
RBAR Implementation: Our optimal strategy consisted of an initial training phase with 50,000 iterations, followed by RBAR-guided domain refinement (illustrated in Fig. 4), which is the 
secondary training phase that has an additional 50,000 training iterations.

Only the RBAR-guided refinement successfully captured the phase transition dynamics. This outcome demonstrates that uniform refinement proves ineffective for solving the dynamic Allen-Cahn equation. The RBAR methodology not only enhances solution accuracy but also helps the PINN framework avoid convergence to incorrect solutions, highlighting its crucial role in capturing complex phase transition dynamics.

\subsection{Simulating complex Morphology with Dynamic Equation}
\label{subsec:complex_morphology}

\begin{figure}[H]
\begin{center}
\includegraphics[width=1\textwidth]{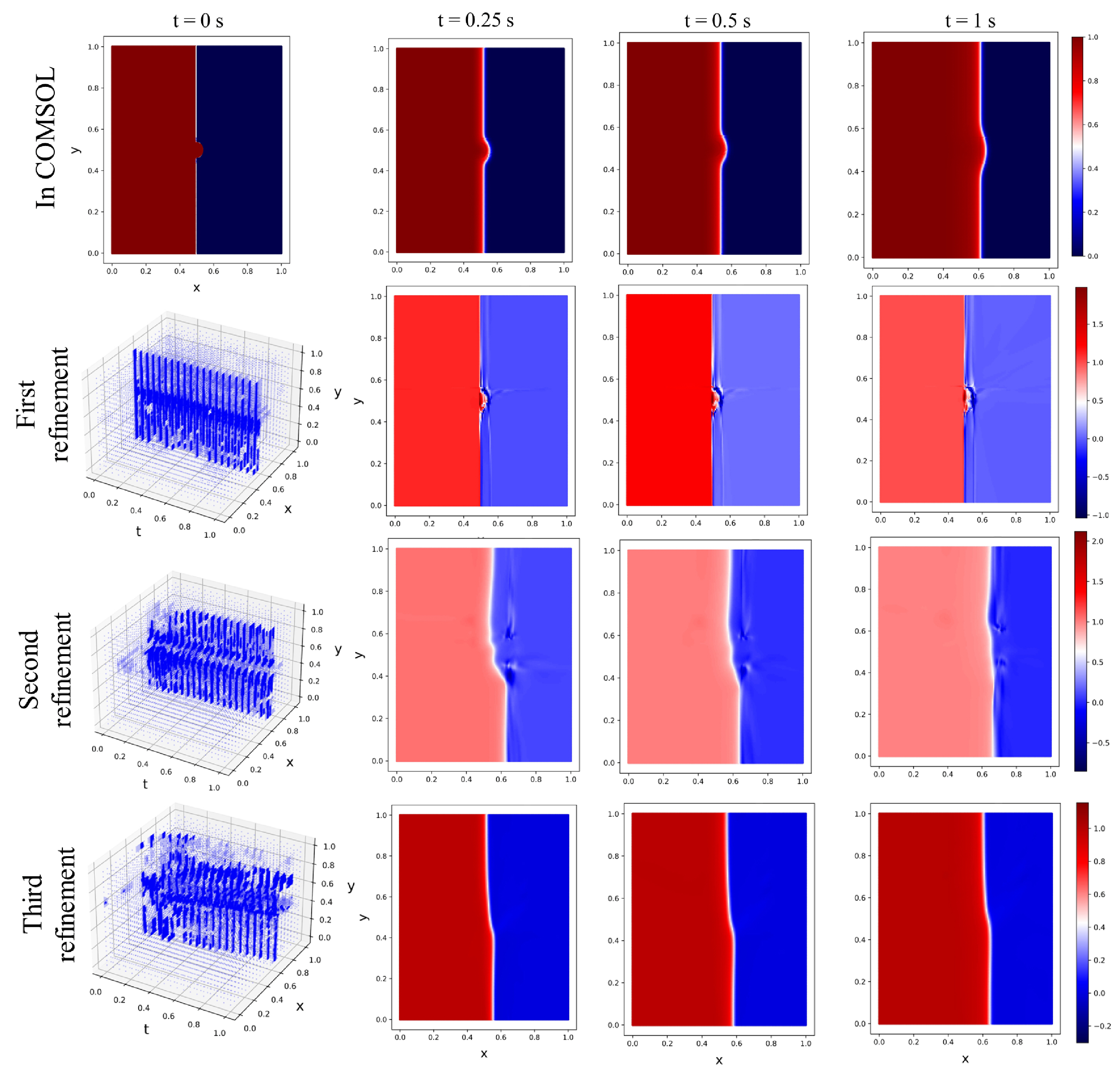}
\caption{Comparative analysis of phase field evolution with initial hump morphology: COMSOL reference solution versus RBAR PINN results after first, second, and third refinement iterations, demonstrating progressive improvement in solution accuracy.}
\label{Fig.4 dynamic hump in RBAR PINN}
\end{center}
\end{figure}

We next investigate the Allen-Cahn equation (Eq. \ref{AC_dynamic_n}) with a more challenging initial condition: a small hump (radius $r = 0.02$) superimposed on the planar interface, as illustrated in Figure \ref{Fig.4 dynamic hump in RBAR PINN}. The evolution of this small perturbation presents a significant computational challenge, even for RBAR-enhanced PINN implementations. 

\noindent \textbf{Limitations of RBAR-Only Implementation:} Initial attempts using only RBAR  revealed significant challenges in accurately capturing the hump evolution. While multiple refinement iterations improved certain aspects of the solution—including better interface smoothness and phase-field values more strictly confined to the $[0,1]$ range—the predicted morphology deviated substantially from the COMSOL reference solution. Specifically, the characteristic hump structure disappeared as the interface evolved toward a planar configuration, indicating fundamental limitations of the RBAR-only approach for this dynamic case.

\begin{figure}[H]
\begin{center}
\includegraphics[width=0.83\textwidth]{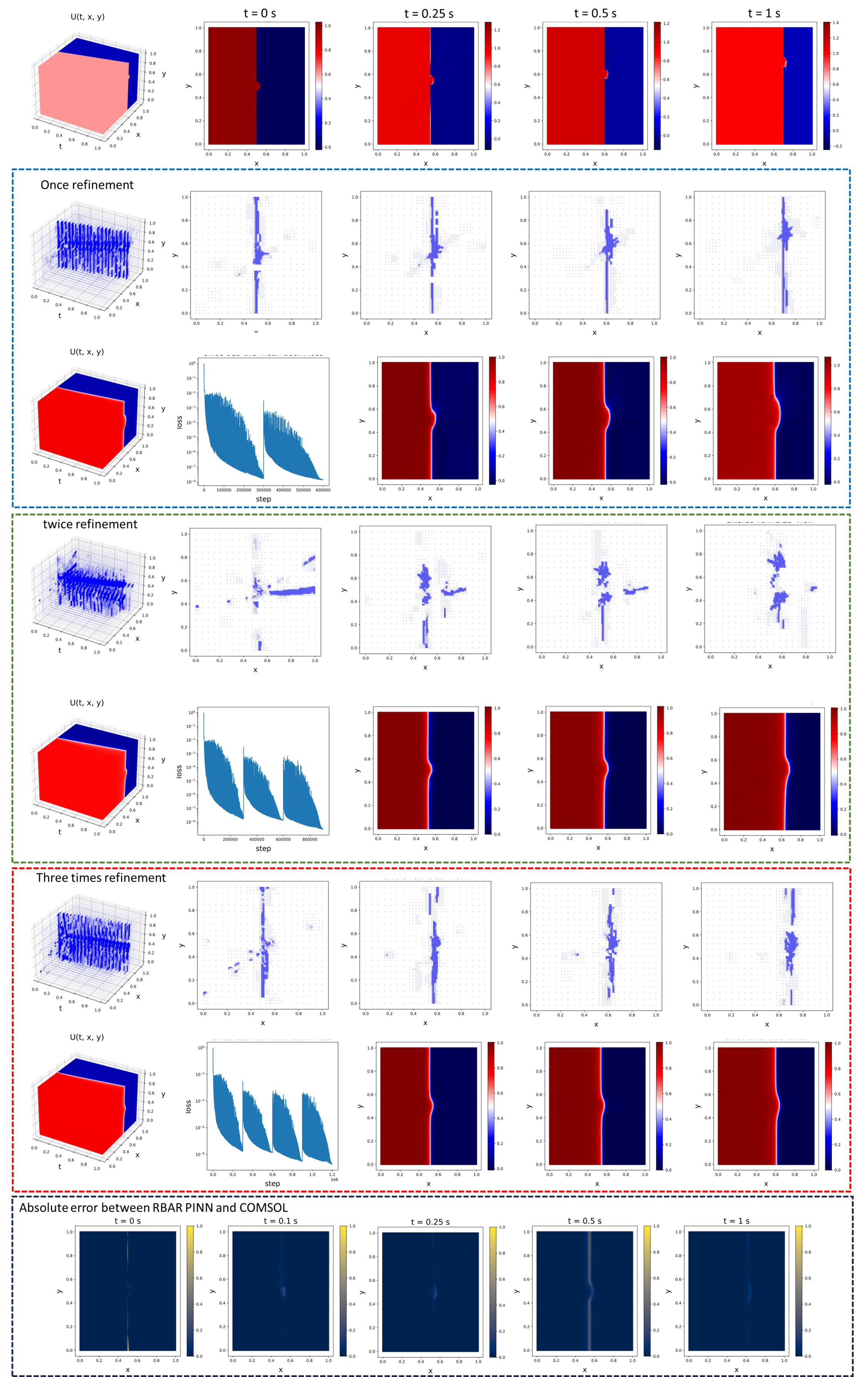}
\caption{Evolution of phase-field solution using RBAR-causality combined PINN methodology: comparison of results with no refinement, first, second, and third refinement iterations, showing progressive improvement in accuracy. The last row presents the absolute error between the three-times-refined PINN predictions and the reference solutions obtained from COMSOL.}
\label{Fig.5_dynamic_hump_in_RBAR-causality_combined_PINN}
\end{center}
\end{figure} 

To address these limitations, we developed a hybrid approach combining RBAR with causality training, as demonstrated in Figure \ref{Fig.5_dynamic_hump_in_RBAR-causality_combined_PINN}. The implementation proceeded through two primary stages: \vspace{-8pt}
\paragraph{Initial Causality Training} The process began with 300,000 iterations using causality-based training. While the results demonstrated correct interface motion direction, they exhibited two significant limitations: (i) inadequate representation of hump growth dynamics, and (ii) substantial spatial overshoot, with interface positions appearing at $x \approx 0.6$ and $x \approx 0.7$ compared to the COMSOL reference positions at $x = 0.5$ and $x = 0.6$, respectively. \vspace{-8pt}

\paragraph{Iterative Refinement Process} Following the initial training, we implemented a systematic refinement procedure comprising: (i) initial refinement based on residual loss distribution, (ii) subsequent 300,000 iterations with causality training, revealing a notable ``overshoot and relocate'' phenomenon that demonstrates the method's adaptive error correction capabilities, and (iii) two additional refinement cycles for solution fine-tuning. This iterative approach progressively improved solution accuracy while maintaining computational efficiency. 

The last row of Figure \ref{Fig.5_dynamic_hump_in_RBAR-causality_combined_PINN} shows the absolute error between the three-times-refined RBAR-PINN predictions and the reference solutions obtained from COMSOL. The error is highly localized at the interface and remains negligible across the rest of the domain.

\begin{figure}[H]
\begin{center}
\includegraphics[width=0.95\textwidth]{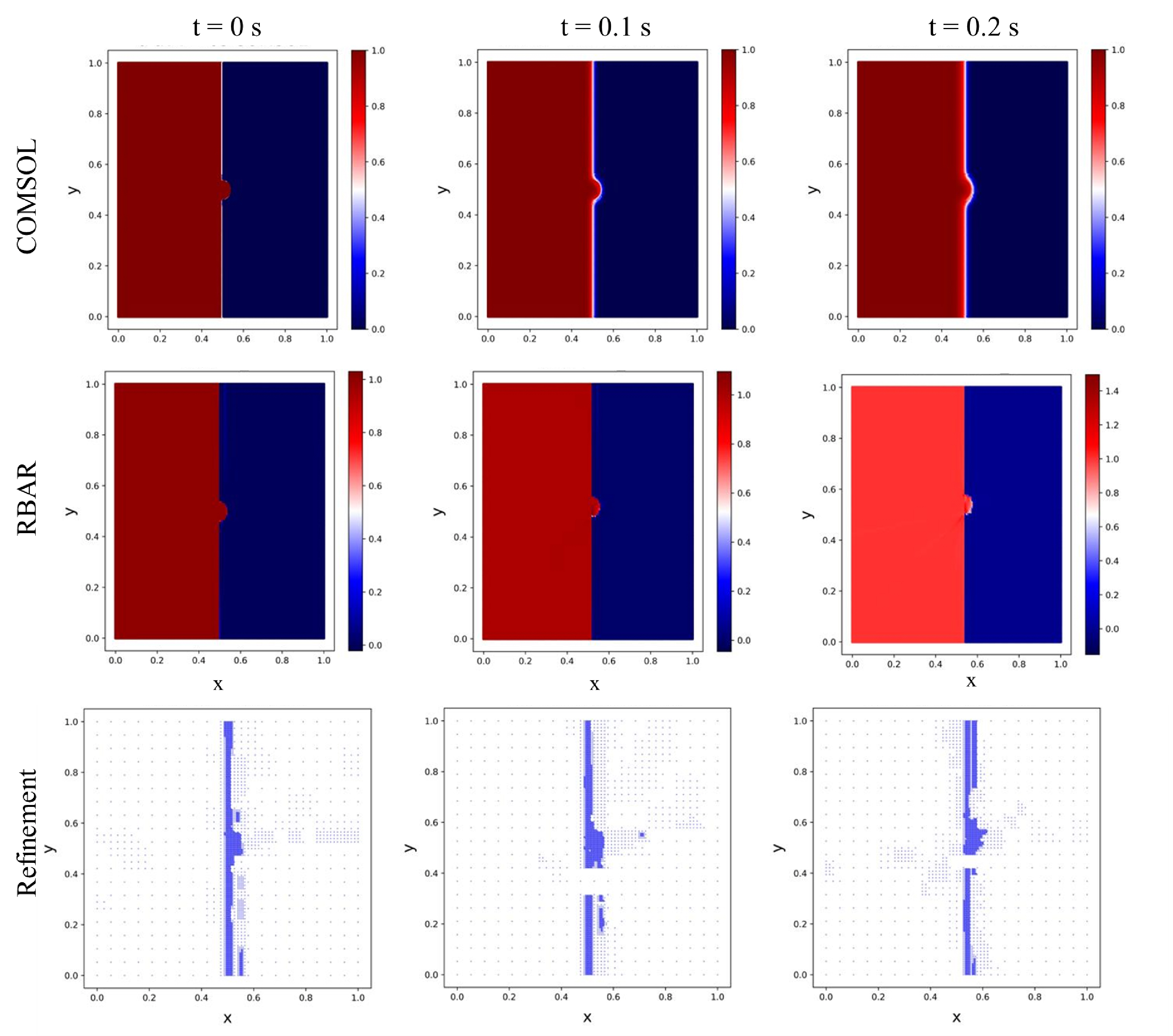}
\caption{Comparative analysis of early-stage evolution (t = 0s, 0.1s, 0.2s) showing phase-field morphology obtained via COMSOL, causality training without refinement, and RBAR patterns.}
\label{Fig.6_RBAR_causality_explain}
\end{center}
\end{figure}

To elucidate the complementary roles of RBAR and causality training, we conducted detailed analysis of the early-time evolution (see Figure \ref{Fig.6_RBAR_causality_explain}). \review{It should be emphasized that the refinement process yields approximately 4,000 points per time step, which is ten times the original count of 400 points.} The causality-training PINN accurately captures the initial morphology dynamics, matching COMSOL results in both interface position and hump structure. RBAR then identifies and refines regions of highest residual loss—primarily the interface and hump regions—providing enhanced spatial resolution where needed most.  This synergistic interaction enables accurate capture of complex morphological features, elimination of spatial overshooting, as well as correct positioning of the evolving interface

The combined methodology successfully reproduces the COMSOL reference solution, demonstrating the effectiveness of integrating RBAR with causality training for challenging phase-field evolution problems. \review{While the well-established finite element software COMSOL demonstrates significantly greater computational efficiency than the RBAR-causality combined PINNs, completing the simulation in approximately 20 seconds compared to 3 hours. It is important to note that physics-informed neural networks are a nascent, mesh-free class of methods for solving PDEs, and they face well-recognized challenges when modeling systems with sharp, evolving interfaces. The objective of the present work is therefore not to compete with mature finite element solvers in terms of raw computational efficiency, but to address a critical failure mode of conventional PINNs. The effectiveness and efficiency of the proposed RBAR–causality combined framework are demonstrated specifically through systematic comparison with traditional PINN formulations, establishing a meaningful advancement for PINN-based modeling of complex interface-evolving systems.}

\section{Conclusion}
\label{sec:summary}

In this study, we introduce a novel methodology that combines RBAR (Residual-Based Adaptive Refinement) with causality-informed training in PINNs to solve spatio-temporal PDEs, with a specific application to the phase field model (PFM) based on Allen-Cahn equations. Our analysis of the dynamic case with planar interfaces reveals that uniform refinement of the computational domain proves inefficient. Instead, the RBAR mechanism enables the PINN to overcome convergence to erroneous solutions by strategically allocating collocation points to capture interface evolution phenomena. However, we observed that RBAR alone proves insufficient for accurately resolving dynamic PFM cases with complex interfacial morphologies. The integration of causality-informed training with RBAR significantly enhances the framework's capability to accurately reproduce temporal interface evolution, particularly in cases involving non-planar morphologies. During the iterative refinement and training process, we identify a characteristic \textit{overshoot and relocate} phenomenon, which emerges from the synergistic interaction between RBAR and causality-informed training. This mechanism operates as follows: initially, RBAR concentrates collocation points in regions of high loss that have been partially resolved through causality training. Subsequently, the causality-informed training utilizes these strategically placed points to obtain accurate solutions at initial timesteps. These refined initial solutions then enable accurate prediction and tracking of interface evolution in subsequent timesteps.

Finally, it should be emphasized that while this study focuses on the benchmark Allen-Cahn equation with fixed parameters, the RBAR-PINN framework is fundamentally designed to be parameter-agnostic. By successfully resolving the numerical``stiffness" associated with sharp, moving interfaces, a known failure point for standard PINN formulations, this work establishes a robust computational foundation for dynamic phase-field modeling. The ability of the algorithm to adaptively redistribute collocation points based on residual ranking ensures that the method can accommodate varying material properties, such as mobility and surface energy, which dictate different interface widths and velocities. Future research will leverage this validated framework to investigate multi-component systems and a broader range of material scales, moving beyond proof-of-concept benchmarks toward complex industrial applications in materials science.

\section*{Author contributions}
\noindent Conceptualization:  WW, SG, HHR \\
Investigation: WW, SG, HHR, TPW\\
Visualization: WW\\
Supervision: SG,  HHR\\
Writing—original draft: WW\\
Writing—review \& editing: WW,  SG,  HHR

\section*{Acknowledgements}
The authors (WW, TPW, and HHR) would like to acknowledge the support by the Hong Kong General Research Fund (GRF) under Grant Numbers 15213619 and 15210622, by an industry collaboration project (HKPolyU Project ID: P0039303), and by a collaborative research fund from HKPolyU (Project ID: P0058676). SG is supported by the U.S. Department of Energy, Office of Science, Office of Advanced Scientific Computing Research grant under Award Number DE-SC0024162 and the 2024 Johns Hopkins Discovery Award. The authors acknowledge the computational resources provided by the University Research Facility in Big Data Analytics (UBDA) at The Hong Kong Polytechnic University.

\section*{Data and code availability}
\noindent All codes and datasets will be made publicly available at {\small\url{https://github.com/Centrum-IntelliPhysics/PINNs-Causality-based-Adaptive-Refinement} upon publication of the work.

\section*{Competing interests}
\noindent The authors declare no competing interest

\bibliographystyle{elsarticle-num}  
\bibliography{ref} 
\end{document}